\documentclass[11pt]{article}

\usepackage{amsmath}

\usepackage{times}
\usepackage{bm}
\usepackage{newtxtext}
\usepackage[subscriptcorrection]{newtxmath}
\usepackage[authoryear]{natbib}
\usepackage[plain,noend]{algorithm2e}

\usepackage{authblk}
\usepackage{graphicx}

\makeatletter
\renewcommand{\algocf@captiontext}[2]{#1\algocf@typo. \AlCapFnt{}#2} 
\def\@algocf@capt@plain{top}
\renewcommand{\algocf@makecaption}[2]{%
  \addtolength{\hsize}{\algomargin}%
  \sbox\@tempboxa{\algocf@captiontext{#1}{#2}}%
  \ifdim\wd\@tempboxa >\hsize
    \hskip .5\algomargin%
    \parbox[t]{\hsize}{\algocf@captiontext{#1}{#2}}
  \else%
    \global\@minipagefalse%
    \hbox to\hsize{\box\@tempboxa}
  \fi%
  \addtolength{\hsize}{-\algomargin}%
}
\makeatother

\usepackage{xcolor}

\newcommand{\nobs}{n_\text{obs}}
\newcommand{\RR}{\mathbb{R}}

\newcommand{\rf}{r}
\newcommand{\EE}{\mathbb{E}}
\newcommand{\PP}{\mathbb{P}}
\newcommand{\bt}{\mathbf{t}}
\newcommand{\bs}{\mathbf{s}}
\newcommand{\bx}{\mathbf{x}}
\newcommand{\bstilde}{\tilde{\mathbf{s}}}
\newcommand{\bxtilde}{\tilde{\mathbf{x}}}

\newcommand{\Var}{\mathrm{Var}}
\newcommand{\Cov}{\mathrm{Cov}}
\newcommand{\Corr}{\mathrm{Corr}}

\newcommand{\dd}{\mathrm{d}}
\newcommand{\one}{\mathbf{1}}

\newtheorem{theorem}{Theorem}

\newtheorem{remark}[theorem]{Remark}
\newtheorem{algo}[theorem]{Algorithm}

\begin{document}

\markboth{M.~Thannheimer, M.~Oesting}{Functional Bayesian inference for Extremes}

\title{Bayesian inference for functional extreme events defined via partially unobserved processes}

\author[1]{Max Thannheimer}
\author[1,2]{Marco Oesting}
\affil[1]{Institute for Stochastics and Applications, University of Stuttgart}
\affil[2]{Stuttgart Center for Simulation Science (SC SimTech),  University of Stuttgart}

\maketitle

\begin{abstract}
In order to describe the extremal behaviour of some stochastic process $X$, approaches from univariate extreme value theory are typically generalized to the spatial domain. In particular, generalized peaks-over-threshold approaches allow for the consideration of single extreme events. These can be flexibly defined as exceedances of a risk functional $\rf$, such as a spatial average, applied to $X$. Inference for the resulting limit process, the so-called $\rf$-Pareto process, requires the evaluation of $\rf(X)$ and thus the knowledge of the whole process $X$. In many practical applications, however, observations of $X$ are only available at scattered sites. To overcome this issue, we propose a two-step MCMC-algorithm in a Bayesian framework. In a first step, we sample from $X$ conditionally on the observations in order to evaluate which observations lead to $\rf$-exceedances. In a second step, we use these exceedances to sample from the posterior distribution of the parameters of the limiting $\rf$-Pareto process. Alternating these steps results in a full Bayesian model for the extremes of $X$. 
We show that, under appropriate assumptions, the probability of classifying an observation as $\rf$-exceedance in the first step converges to the desired probability. Furthermore, given the first step, the distribution of the Markov chain constructed in the second step converges to the posterior distribution of interest. The procedure is compared to the Bayesian version of the standard procedure in a simulation study. 
\end{abstract}


\section{Introduction}

Modelling extreme events in the framework of extreme value theory is an important problem in modern statistics which has been studied extensively. For environmental applications such as rainfall events, complex extreme events have to be described, interfered with, and simulated in a spatial or spatiotemporal setting. 

Two main approaches are available for statistical inference in extreme value theory. The block maxima approach proposes to divide data consisting of repeated measurements into blocks, often yearly, and to consider the maximum within each block only. In case of spatial or spatiotemporal data, maxima are taken pointwise and, thus, the practice of taking yearly maxima does not only heavily thin out the data, but might also result in the conflation of different extreme events. This conflation typically also leads to complex likelihoods in statistical models for spatial maxima, such as max-stable processes, rendering inference and simulation challenging, particularly in higher dimensions, {see, for instance, \cite{stephensontawn05}, \cite{DEO17} and \cite{HDRG19}.}

The second main approach, the peaks-over-threshold approach, allows to focus on single extreme events defined as exceedances over a high threshold. While the definition of exceedances is straightforward in a univariate framework, the spatial setting requires the introduction of a so-called risk functional $r$, which maps the process realization to a scalar value. 

The choice of the risk functional $r$ is given by the application. In principle, different risk functionals can select very different events as exceedances to be modelled. For instance, there are typically two types of heavy precipitation events that may lead to flooding. Either cyclonic events, with a large spatial range but overall lower intensity, or convective events, with a very high intensity but only small spatial extent. \cite{coles-tawn-1996} studied extreme areal rainfall via exceedances of the spatial integral, which correspond to extreme cyclonic events. {In contrast, the definition of extremes via the spatial supremum as suggested in \cite{ferreira-dehaan-2014} rather corresponds} to extreme convective events. For a data set of precipitation in Zurich, \cite{defondeville-davison-2018} demonstrated that the largest convective and cyclonic events, also defined via the spatial integral and an approximation of the supremum, do not share any common event. Therefore, depending on the risk functional completely different events might be chosen. For possible mixtures of various physical processes, the choice of the risk functional allows for the focus on the component of interest. 

{Generalizing the construction of Pareto processes by \cite{ferreira-dehaan-2014} based on the spatial supremum,} \cite{dombry-ribatet-2015} introduced the concept of general $r$-Pareto processes, the spatial limit process for the peaks over threshold approach, for arbitrary nonnegative homogeneous risk functionals $r$, showing their connection to the corresponding max-stable processes and explicitly calculating their likelihoods {in the case of marginal distributions following asymptotically a power law decay. More generally,} \cite{defondeville-davison-2022} introduced generalized $r$-Pareto processes capable of handling a broader class of possible risk functionals $r$, all types of tail decay and arbitrary margins, leading to more complex formulas and marginal transformations. 

{In contrast to the family of generalized Pareto distributions in the univariate} setting, the {family of spatial $r$-Pareto processes is infinite-dimensional and, consequently, inference is often performed for finite-dimensional parametric subfamilies.}
{Apart from few exceptions such as the recent likelihood-free approach using neural Bayes estimators by \cite{Richards-et-al-2024}, most of parametric inference approaches for $r$-Pareto processes so far are likelihood-based and, to the best our knowledge, have been studied in a frequentist setting. However,} inference {based on} the full likelihood becomes quickly intractable in higher dimensions, which are not avoidable in a spatial setting. {For the special case of the Brown--Resnick process,} \cite{engelke-et-al-2015} introduce a spectral estimator. \cite{defondeville-davison-2018} compare this approach with a composite likelihood approach based on \cite{wadsworth-tawn-2014} and their own score matching-based one.

{In this paper, we go in a different direction and} follow a full Bayesian approach giving us access to the full posterior distribution and all the related Bayesian techniques. Generating samples from the fitted model comes naturally with our method and respects uncertainty in the fit. 

Furthermore, our approach makes use of conditional sampling which allows us to choose any homogeneous risk functional {relying on values of the process} on an arbitrarily fine grid, even when our observed data is sparse and irregular. One application could be rainfall modeling, where {the network of weather stations providing highly reliable observation} data is {typically very} sparse, but rainfall as a process is highly complex and spatially {varying at} high resolution. When interested in potential flooding one needs a complex risk functional which cannot be approximated via just the low dimensional sparse weather data {set}. Compared to other methods, {our approach does not require an a priori choice of a} coarse approximation of the risk functional. {In contrast, as we will demonstrate} in a simulation study, using the real risk functional instead of a approximation {leads to significant gains} in the estimation {accuracy}.

\section{Mathematical Foundations}
Let $S \subset \mathbb{R}^d$ be compact and $C_+(S)$  be the space of non-negative continuous real-valued functions $f: S \to [0,\infty)$ equipped with norm $\|f\|_\infty := \sup_{s \in S} |f(S)|$ {and the $\sigma$-algebra generated by cylinder sets}. Consider a sample-continuous process $X=\{X(s): \, s \in S\}$ which is in the max-domain of attraction of some sample-continuous max-stable process $Z=\{Z(s): \, s \in S\}$ with unit scale $\alpha$-Fr\'echet margins for some $\alpha>0$, i.e.\ there exist continuous functions $a_n: S \to (0,\infty)$ such that
$$ \left\{ \max_{i=1}^n \frac{X_i(s)}{a_n(s)}, \, s \in S\right\} \to_d \{Z(s), s \in S\} $$
weakly in $C_+(S)$ for independent stochastic processes $X_1, X_2, \ldots$ with the same law as $X$.
\medskip

By \cite{{dehaan-84}}, the process $Z$ allows for the spectral representation
\begin{align}
    Z(s) = \bigvee_{k=1}^\infty \Gamma_k^{-1/\alpha} W_k(s), \quad s \in S,
    \label{spectral_representation}
\end{align}  
where $\{\Gamma_k\}_{k \in \mathbb{N}}$ are the arrival times of a Poisson process on $(0,\infty)$ with unit intensity and $W_k$ are independent copies of a non-negative sample-continuous stochastic process 
$W = \{W(s): \, s \in S\}$, the so-called \textit{spectral process}, satisfying 
$\EE(W(s)^\alpha)=1$ for all $s \in S$. Further, sample-continuity of $Z$ also implies that $\EE(\sup_{s\in S}W(s)^\alpha)<\infty$, compare Corollary 9.4.5 in \cite{dehaan-ferreira-2006}. It is important to note that the law of the spectral process~$W$ is not uniquely determined by the resulting max-stable process~$Z$. Henceforth, we will consider a choice that is standardized with respect to some reference point $s_0 \in S$. Provided that there is a choice of the spectral process $W$ that satisfies $W(s_0) > 0$ almost surely, without loss of generality, this process can be chosen such that $W(s_0)=1$ almost surely, see, for instance, Theorem 5 in \cite{penrose-1992}. The law of this choice is unique.
\medskip

\subsection{$\rf$-Normalized Spectral Representation and $\rf$-Pareto Processes} Let $\rf: C_+(S) \to \mathbb{R}$ be a non-trivial, homogeneous, continuous functional, the so-called \textit{risk functional}  \citep{defondeville-davison-2018}. Provided that $\rf(W) > 0$ almost surely, there exists a non-negative sample-continuous stochastic process $W^{(\rf)} = \{W^{(\rf)}(s), \, s \in S\}$ such that the equality
\begin{align*}
Z(s) =_d  \sqrt[\alpha]{c_{\rf}(\alpha)} \bigvee_{k=1}^\infty \Gamma_k^{-1/\alpha} \frac{W^{(\rf)}_k(s)}{\rf(W^{(\rf)}_k)}, \quad s \in S,  
\end{align*} 
holds in distribution where
$W^{(\rf)}_1, W^{(\rf)}_2, \ldots$ are independent copies of $W^{(\rf)}$ and
\begin{align*}
c_{\rf}(\alpha) = \int_{C_+(S)} \rf (w)^\alpha \, \mathrm{d} \PP_W(w).
\end{align*}
The law of the standardized process ${W^{(\rf)}}/{\rf \left( W^{(\rf)}\right)}$ on the unit pseudo-sphere $$\mathbb{S}^{(\rf)}:= \left\{f\in C_+(S) : \, \rf(f)=1\right\}$$ 
with respect to $\rf$ is uniquely given by
\begin{align*}
\PP\left(\frac{W^{(\rf)}}{\rf\left(W^{(\rf)}\right)} \in A\right) = \frac{1}{c_{\rf}(\alpha)} \int_{C_+(S)} \rf(w)^\alpha \one{\left\{\frac{w}{\rf(w)} \in A\right\}} \,\mathrm{d} \PP_W (w),
\quad A \subset \mathbb{S}^{(\rf)} \text{ measurable},
\end{align*}
see \cite{dombry-ribatet-2015}.
This is, for instance, satisfied by the process $W^{(\rf)}$ {whose law is} given by
\begin{align}
\PP\left(W^{(\rf)} \in B \right) = \frac{1}{c_{\rf}(\alpha)} \int_{C_+(S)} \rf(w)^\alpha \one{\left\{w \in B\right\}} \, \mathrm{d} \PP_W (w), \quad B\in \mathcal{B}\left( C_+(S)\right) .
\label{W^ell}
\end{align}
This measure transform can be interpreted as reweighting of the distribution of $W$ while keeping the actual range of the sample paths unchanged. In particular, $W^{(\rf)}(s_0)=1$ almost surely if $W(s_0)=1$ almost surely.

We define extreme events as risk functional exceedances, i.e., realizations of $X$ with a risk functional evaluation above a high threshold $u$. More precisely, we call a realization of $X$ an $\rf$-exceedance over a threshold $u$ if $\rf(X) > u$.
For the process $X$  in the max-domain of attraction of the max-stable process $Z$ the weak convergence 
\begin{align*}
\lim_{u\to \infty} \PP \left(\frac{X}{u} \in \cdot \,\bigg|\, \rf(X) > u \right) = \PP\left(P^{(\rf)}\in \cdot \right) ,
\end{align*}
holds in $C_+(S)$ with the limit process $P^{(\rf)}$ being the so-called $\rf$\textit{-Pareto process} \citep[][Theorem 3]{dombry-ribatet-2015}.
By decomposing $X$ into a $\rf$-normalized part ${X} / {\rf(X)}$ and an intensity part $\rf(X)$ we obtain
\begin{align}
\lim_{u\to \infty} \PP  \left( \rf(X)>t u, \frac{X}{{\rf}(X)} \in A \,\bigg|\, \rf(X) > u \right) = t^{-\alpha} \cdot \PP\left(\frac{W^{(\rf)}}{\rf(W^{(\rf)})}\in A\right) , \quad t>1,\ A\in \mathbb{S}^{(\rf)},
\label{ellParetoLimit}
\end{align}
i.e.\ we can decompose the $\rf$-Pareto process via 
\begin{equation*}
P^{(\rf)}=_d P_\alpha \cdot  \frac{W^{(\rf)}}{\rf(W^{(\rf)})}, 
\end{equation*}
where $ P_\alpha$ is {an} $\alpha$-Pareto random variable independent from $W^{(\rf)}$.

\subsection{Example of Spectral Functions: Brown-Resnick Process }
\label{subsec:BR}
A {popular} choice for a max-stable process is the so-called \textit{Brown-Resnick process}. For the spectral process $W$ in \eqref{spectral_representation} normalized with respect to $s_0$ we have the representation
\begin{align}
 W(s) = \exp\left( \frac 1 \alpha \left[G(s)-G(s_0) - \gamma(s-s_0) \right]\right), \quad s \in \RR^d,
\label{def:BR}
\end{align}
where $G$ is a centered Gaussian process with stationary increments and semi-variogram 
$$ \gamma(h) = \frac 1 2 \Var( G(h) - G(0) ), \quad h \in \RR^d. $$
The spectral process only depends on the semi-variogram and
the resulting max-stable process is stationary \citep{KSH-2009}.

For statistical inference, one typically introduces some parameterization for the semi-variogram $\gamma$. A convenient choice is for instance the \textit{power variogram}
\begin{align}
\gamma(h)=c \|h\|^\beta, \quad h \in \RR^d,
\label{def:powervariogram}
\end{align}

with $\beta\in (0,2]$ and $c>0$. One advantage in choosing {the} log-Gaussian structure is that both conditional and unconditional simulation from these processes are well-studied problems.

\section{Bayesian Inference in Case of Observable Risk Functional}
\label{sec:risk-observable}

Henceforth, we will develop procedures for statistical inference for the extremal behavior of the stochastic process $X$ (defined as $r$-exceedances) based on observations at sites $s_0, s_1, \ldots, s_n$.
To simplify notation, but still stress the specific role of $s_0 \in \RR$ as normalizing site, for any random or deterministic function $f: S \to \RR$, we write 
$f(\bs) = ( f(s_1), \ldots, f(s_n))$, for short.

In this section, we assume that the risk functional $\rf$ is observable, i.e., $\rf(X)$ depends on the observed values $X(s_0)$ and $X(\bs)$ only and with some slight abuse of notation, we may write
$$ \rf(X) =: \rf((X(s_0), X(s_1), \ldots, X(s_n))).$$
To further simplify notation, we define
$$ \rf(f(\bs)) := \rf((1, f(\bs))). $$
In other words, $\rf$ applied to an $n$-dimensional vector $\bx$ will be 
interpreted as $\rf$ applied to the $(n+1)$-dimensional vector $(1, \bx)$. This short notation will turn out to be particularly useful when evaluating $\rf(W)$ or 
$\rf(W^{(\rf)})$ which simplify to $\rf(W(\bs))$ and
$\rf(W^{(\rf)}(\bs))$ as $W(s_0) = W^{(\rf)}(s_0)=1$ almost surely.

We propose an inference procedure  that is based on the limit theorem \eqref{ellParetoLimit}, but considers the two factors 
$P_\alpha/\rf(W^{(\rf)})$ and $W^{(\rf)}$ of the limiting $\rf$-Pareto process $P^{(\rf)}$ separately. More precisely,
\begin{align}
 \lim_{u\to \infty} \PP \left( \bigg(\frac{X(s_0)}{u} ,\frac{X(\bs)}{X(s_0)}\bigg) \in \cdot \,\bigg|\, \rf(X) > u \right)
=& \lim_{u\to \infty} \PP \left(  \left(u^{-1} X(s_0), \frac{u^{-1} X(\bs)}{u^{-1} X(s_0)} \right) \in \cdot \,\bigg|\, \rf(X) > u \right) \nonumber \\
=& \PP\left( \bigg(\frac{P_\alpha }{\rf \left( W^{(\rf)}(\bs)\right)} {,} W^{(\rf)}(\bs)\bigg) \in \cdot \right),
    \label{LimitLawForInference}
\end{align}
where we use that $W^{(\rf)}(s_0)=1$ almost surely.
For suitably large risk functional evaluations $\rf(X)>u$, we assume the limit relation to hold and therefore have
\begin{align}\label{LimitLawForInference2}
    \left(\frac{X(s_0)}{u}, \frac{X(\bs)}{X(s_0)}\right) =_d \left(\frac{P_\alpha}{\rf \left( W^{(\rf)}(\bs)\right)}, W^{(\rf)}(\bs)\right).
\end{align}
in distribution as a basis for statistical inference. If $X$ is an $\rf$-Pareto process, Equations \eqref{LimitLawForInference} and \eqref{LimitLawForInference2} do not only hold asymptotically conditioning on an $\rf$-exceedance over high thresholds, but also unconditionally.

Statistical inference on the the distribution of $W^{(\rf)}$ will be done in a parametric Bayesian framework. To this end, besides considering the parameter $\alpha \in (0,\infty)$ of the Pareto random variable $P_\alpha$, we also introduce a parametrization of the distribution of $W$  with parameter 
$\theta \in \Theta \subset \mathbb{R}^k$ and a prior density $\pi(\theta,\alpha)$ on the parameter space $\Theta \times (0,\infty)$. Given an observation $(x(s_0),x(\bs))$ that belongs to an $\rf$-exceedance over the threshold $u$, i.e., $r(x)>u$, we get the joint posterior distribution 
\begin{align}
    &f(\theta,\alpha \mid x(s_0), \dots, x(s_n)) \nonumber  \\
    &\propto
    f_{(X(s_0)/u,X(\bs)/u)}\left(\frac{x(s_0)}{u}, \dots, \frac{x(s_n)}{u} \,\middle  \vert \, \theta, \alpha\right)\pi(\theta,\alpha) \nonumber  \\ 
    &\propto
    f_{(X(s_0) / u, X(\bs)/ X(s_0))}\left(\frac{x(s_0)}{u}, \frac{x(\bs)}{x(s_0)} \, \middle \vert \, \theta, \alpha\right)\pi(\theta,\alpha) \nonumber  \\
    &=
    f_{(X(s_0)/ u\mid X(\bs)/ X(s_0))}\left(\frac{x(s_0)}{u} \,\middle \vert \,  \frac{x(\bs)}{ x(s_0)}, \theta, \alpha\right) \cdot f_{X(\bs)/ X(s_0)}\left(\frac{x(\bs)}{x(s_0)} \,\middle \vert \, \theta,\alpha\right) \cdot \pi(\theta,\alpha) \nonumber  \\
    &=f_{\left(P_\alpha/\rf\left(W^{(\rf)}(\bs)\right) \middle \vert W^{(\rf)}(\bs))\right)}\left(\frac{x(s_0)}{u} \, \middle \vert \,  \frac{x(\bs)}{ x(s_0)}, \theta, \alpha\right) \cdot f_{W^{(\rf)}(\bs)}\left(\frac{x(\bs)}{ x(s_0)} \,\middle \vert \, \theta,\alpha \right) \cdot \pi(\theta,\alpha)
    \label{posterior_unobserved_rf_single_observation_1}
\end{align}
where, in the last equality, assumption  \eqref{LimitLawForInference2} is used. Firstly, we derive a density for the distribution of $P_\alpha / \rf \left( W^{(\rf)} \right)$ conditional on $W^{(\rf)}(\bs)$. It holds 
\begin{align*}
    \PP\left( \frac{P_\alpha }{\rf \left( W^{(\rf)}(\bs)\right)}  \leq x \, \middle \vert  \, W^{(\rf)}(\bs)=w \right) &= \PP\left( \frac{P_\alpha }{\rf \left( w\right)}  \leq x \, \middle \vert  \, W^{(\rf)}(\bs)=w \right) \\
    &=\PP\left( P_\alpha \leq x \rf(w) \right) 
    ={1 -}\one\{x\rf(w)>1\} \cdot \left( x \rf(w) \right)^{-\alpha}.
\end{align*}
Differentiation leads to
\begin{align}
    f_{{P_\alpha / \rf \left( W^{(\rf)}(\bs)\right)}}(x \mid  W^{(\rf)}(\bs)=w )
    &=\one\{x\rf(w)>1\} \rf(w)^{-\alpha} \alpha x^{-\alpha-1}.
    \label{intensity density}
\end{align}

For the second factor of the right-hand side of \eqref{posterior_unobserved_rf_single_observation_1}, we note that, when a Lebesgue density of the $n$-dimensional random vector $W(\bs)$ exists, the measure transform  \eqref{W^ell} of $W^{(\rf)}$ yields
\begin{align}
    f_{W^{(\rf)}(\mathbf{s})}(w(\bs)\mid \alpha)= \frac{1}{c_{\rf}(\alpha)} \rf(w(\bs))^\alpha f_{W(\bs)}(w(\bs)\mid \alpha).
    \label{spectral density}
\end{align}
Combining both densities, \eqref{intensity density} and \eqref{spectral density}, leads to
 \begin{align}
    &f(\theta,\alpha \mid x(s_0), \dots, x(s_n)) \nonumber  \\
    &\propto
    f_{\left(P_\alpha/\rf\left(W^{(\rf)}(\bs)\right) \middle \vert W^{(\rf)}(\bs))\right)}(x(s_0)/u \mid  x(\bs)/ x(s_0), \theta, \alpha) \cdot f_{W^{(\rf)}(\bs)}(x(\bs)/ x(s_0) \mid \theta,\alpha) \cdot \pi(\theta,\alpha)\nonumber \\
    &=\one\{\rf(x(\bs))>u\} \rf(x(\bs)/ x(s_0))^{-\alpha} \alpha \left(x(s_0)/u\right)^{-\alpha-1} \nonumber \\
    & \qquad \cdot \frac{1}{c_{\rf}(\alpha,\theta)} \rf(x(\bs)/ x(s_0))^\alpha f_{W(\bs)}(x(\bs)/ x(s_0)\mid \theta,\alpha) \cdot \pi(\theta,\alpha) \nonumber \\
    &= \one\{\rf(x(\bs))>u\} \cdot \alpha \left(x(s_0)/u\right)^{-\alpha-1}  \cdot \frac{1}{c_{\rf}(\alpha,\theta)} f_{W(\bs)}(x(\bs)/ x(s_0)\mid \theta,\alpha) \cdot \pi(\theta,\alpha).
    \label{posterior_unobserved_rf_single_observation_2}
\end{align}

For observations $(x_1(s_0), x_1(\bs)), \ldots, (x_m(s_0), x_m(\bs))$, accompanied by the corresponding risk functional evaluations $\rf(x_1), \ldots, \rf(x_m)$, for realizations $x_1,\ldots,x_m$ from $m$ independent copies of $X$, we define the index set of risk functional exceedances over the threshold $u$ as 
\begin{align}
I(u):=\left\{ i\in \left\{ 1, \dots, m \right\}  : \, \rf\left(x_i\right)>u \right\}.
    \label{set_of_exceedances}
\end{align}
Analogously to Equation \eqref{posterior_unobserved_rf_single_observation_2}, 
the posterior distribution can be written as product density
 \begin{align}
    &f\left(\theta,\alpha \, \middle \vert \, \left\{x_i(s_0), x_i(\bs) \right\}_{i \in I(u)} \right) \nonumber  \\
    &\propto  \pi(\theta,\alpha) \cdot  \prod_{{i=1}}^{{m}} \one\{\rf(x_i(\bs)>u\} \cdot \alpha \left(x_i(s_0)/u\right)^{-\alpha-1}  \cdot \frac{1}{c_{\rf}(\alpha,\theta)} f_{W(\bs)}(x_i(\bs)/ x_i(s_0)\mid \theta,\alpha) \nonumber  \\
    &=\pi(\theta,\alpha)  \cdot \alpha^{|I(u)|} c_{\rf}(\alpha,\theta)^{-|I(u)|} \prod_{i\in I(u)} \left(x_i(s_0)/u\right)^{-\alpha-1}  f_{W(\bs)}(x_i(\bs)/ x_i(s_0)\mid \theta,\alpha) \nonumber \\
    &=:\pi(\theta,\alpha)  \cdot g\left(\left\{x_i(s_0), x_i(\bs) \right\}_{i \in I(u)} \, \middle \vert \,\theta,\alpha\right)
    \label{posterior_unobserved_rf_multiple_observations}
\end{align}
Sampling from this posterior distribution can be performed via a Metropolis--Hastings approach for an iterative algorithm. Given some current parameter $(\theta,\alpha) \in \Theta \times (0,\infty)$, a new parameter $\left(\theta^\prime,\alpha^\prime\right)$ is proposed from a suitable proposal distribution $q((\theta,\alpha),(\cdot,\cdot))$. Then we accept the proposal $(\theta^\prime,\alpha^\prime)$ with probability 
\begin{align*}
    a((\theta,\alpha),(\theta^\prime,\alpha^\prime))
    =\min\left\{ \frac{ q((\theta^\prime,\alpha^\prime),(\theta,\alpha))  \pi (\theta^\prime,\alpha^\prime)  g\left(\left\{x_i(s_0), x_i(\bs) \right\}_{i \in I(u)} \, \middle \vert \,\theta^\prime,\alpha^\prime\right)}{q((\theta,\alpha),(\theta^\prime,\alpha^\prime)) \pi (\theta,\alpha)  g\left(\left\{x_i(s_0), x_i(\bs) \right\}_{i \in I(u)} \, \middle \vert \,\theta,\alpha\right) },1 \right\}
\end{align*}
For Metropolis kernels, it holds (compare Theorem 1 and Corollary 2 in \citealp{tierney-1994}) that the invariant distribution of the resulting Markov chain is unique and convergence towards it is guaranteed in total variation norm, if the proposal is chosen in such a way that irreducibility and aperiodicity of the Markov chain are guaranteed. The irreducibility holds if we can reach the whole parameter space $\Theta \times (0,\infty)$ within a finite number of steps with positive probability. Aperiodicity follows if, additionally, the algorithm can stay in the same state with positive probability. Both properties are satisfied with standard choices of the proposal like appropriate independent samplers or random walk samplers. 
Thus, we have shown that the Markov chain constructed via the proposed Metropolis-Hastings algorithm will converge towards the true posterior distribution \eqref{posterior_unobserved_rf_multiple_observations}. More precisely, the following theorem holds. 

\begin{theorem}
If $X$ is an $\rf$-Pareto process and the proposal is chosen in such a way that irreducibility and aperiodicity are guaranteed, the output chain of the MCMC algorithm $\left(\theta^{(n)},\alpha^{(n)} \right)_{n \in \mathbb{N}}$ converges to the posterior in total variation norm, i.e.,
\begin{align*}
\lim_{n \to \infty}\left\|  \PP\left( (\theta^{(n)},\alpha^{(n)}) \in \, \cdot  \right) - \PP\left( (\theta,\alpha) \in \, \cdot \,\middle \vert \, \left\{x_i(s_0), x_i(\bs) \right\}_{i \in I(u)}\right) \right\|_{\text{TV}} = 0 , 
\end{align*}
for arbitrary starting values $\theta^{(0)} \in \Theta,\alpha^{(0)} > 0$.

\end{theorem}

\section{Bayesian Inference in Case of Unobservable Risk Functional}
\label{sec:risk-unobservable}
In this section, we assume the risk functional is unobservable, i.e., the evaluation of the risk functional $r(f):= r(f(t_1),\dots,f(t_N))$ requires values on a finer set of sites $\mathbf{t}:= ( t_1, \dots, t_N)$ with $n\ll N$, but we have only observations of $X(s_0)$ and $X(\bs):=(X(s_1),\dots, X(s_n))$ at some sites $s_0,\dots,s_n \in S$ available. Therefore, we need the values $X(\bt):= (X(t_1),\dots,X(t_N))$ conditional on the available observations of $X(s_0)$ and $X(\bs)$. For more general risk functionals, we assume that the evaluation at a large number of sites $N$ gives an appropriate approximation and therefore this restriction is acceptable.

\subsection{Conditional Fine Grid Sampling}
\label{subsec:cond fine grid sampling}
{As the exact value of $r(X)$ is not uniquely determined by the observations, we will  need to make further assumptions that allow for inference. Here, rather than assuming the validity of \eqref{LimitLawForInference} for the joint distribution of $X(s_0)$ and $X(\bs)/X(s_0)$, which holds conditional on an exceedance over some sufficiently high threshold (only), we assume the equality 
\begin{align} \label{LimitLawForInference3}
  W^{(\rf)}(\bs) =_d   \frac{X(\bs)}{X(s_0)} 
\end{align}
in distribution even unconditionally. Note again that \eqref{LimitLawForInference} and, consequently, also \eqref{LimitLawForInference3} are satisfied if, for instance, $X$ is an $\rf$-Pareto process. However, \eqref{LimitLawForInference3} is much weaker as it does not impose any restrictions on the marginal distribution of $X(s_0)$.}

Firstly, we sample from
\begin{align*}
   W^{(\rf)} &\mid W^{(\rf)}(s_0)=\frac{x(s_0)}{x(s_0)},W^{(\rf)}(\bs)=\frac{x(\bs)}{x(s_0)}\\
   W^{(\rf)} & \mid W^{(\rf)}(\bs)=\frac{x(\bs)}{x(s_0)}. 
\end{align*}
Therefore we take a look at the conditional distributions of $W^{(\rf)}$.
The joint distribution on the observations and the finer grid fulfils
\begin{align*}
&   \PP \left( W^{(\rf)}(\bt) \in A ,W^{(\rf)}(\bs) \in B \right) \\
 =    &\frac{1}{c_{\rf}(\alpha)} \int_{C_+(S)} \rf(\mathbf{z})^\alpha \one{\left\{ \mathbf{z} \in A \right\}} \one{\left\{ \mathbf{x} \in B \right\}}  \PP_{(W(\bt), W(\bs))}(\dd (\mathbf{z},\mathbf{x})) \\
= & \frac{1}{c_{\rf}(\alpha)} \int_{B}\int_{A} \rf(\mathbf{z})^\alpha   \PP_{W(\bt) \mid W(\bs)=\mathbf{x} } \left( \dd (\mathbf{z}) \right)\PP_{ W(\bs)}  \left(\dd (\mathbf{x})\right) 
\end{align*}
and therefore the conditional density of $W^{(\rf)}(\bt) \mid W^{(\rf)}(\bstilde)=\tilde{\mathbf{x}} $ 
\begin{align*}
   f_{W^{(\rf)}(\bt) \mid W^{(\rf)}(\bstilde)=\tilde{\mathbf{x}}}(\mathbf{z}\mid \theta, \alpha)&=\frac{f_{(W^{(\rf)}(\bt), W^{(\rf)}(\bstilde))}(\mathbf{z},\tilde{\mathbf{x}}\mid \theta, \alpha)}{\int_{\RR^{N}}f_{(W^{(\rf)}(\bt), W^{(\rf)}(\bstilde))}(\tilde{\mathbf{z}},\tilde{\mathbf{x}}\mid \theta, \alpha) \dd \tilde{\mathbf{z}}}\\
   &=\frac{1}{c_{\rf}(\alpha)} {\rf}^\alpha (\mathbf{z} ) f_{W(\bt) \mid W(\bstilde)=\tilde{\mathbf{x}} } (\mathbf{z}\mid \theta, \alpha) \frac{f_{W(\bstilde)}(\tilde{\mathbf{x}}\mid \theta, \alpha)}{\int_{\RR^{N}}f_{(W^{(\rf)}(\bt), W^{(\rf)}(\bstilde))}(\tilde{\mathbf{z}},\tilde{\mathbf{x}}\mid \theta, \alpha) \dd \tilde{\mathbf{z}}}\\
  &=:\frac{1}{c_{\rf}(\alpha)} C(\bxtilde \, \vert \, \theta, \alpha)  {\rf}^\alpha (\mathbf{z} ) f_{W(\bt) \mid W(\bstilde)=\tilde{\mathbf{x}} } (\mathbf{z}\mid \theta, \alpha)  , \quad \mathbf{z} \in \RR^N.
\end{align*}
We choose $\tilde{\mathbf{x}} $ as justified by the assumption  \eqref{LimitLawForInference} as  observations
$$\tilde{\mathbf{x}}= \frac{x(\bs)}{x(s_0)}.$$ 

Now that we know the conditional density, we can construct a Metropolis-Hastings algorithm for the conditional simulation. As proposal density $q(\cdot,\cdot)$ we choose the conditional density of the spectral process $W(\bt) \mid W(\mathbf{s})=\tilde{\mathbf{x}}  $ itself, i.e., independently from the current realization $\bf{w}$ we choose a new $\mathbf{w}^\prime$ according to
\begin{align}
    q(\mathbf{w},\mathbf{w}^\prime)=f_{W(\bt) \mid W(\mathbf{s})=\tilde{\mathbf{x}} } (\mathbf{w}^\prime\mid \theta, \alpha).
\end{align}
Using this proposal, we get 
\begin{align*}
    a(\mathbf{w},\mathbf{w}^\prime )&=\min \left\{ \frac{q(\mathbf{w}^\prime,\mathbf{w}) \cdot f_{W^{(\rf)}(\bt) \mid W^{(\rf)}(\mathbf{s})=\tilde{\mathbf{x}}}(\mathbf{w}^\prime \mid \theta, \alpha)}{q(\mathbf{w},\mathbf{w}^\prime)\cdot f_{W^{(\rf)}(\bt) \mid W^{(\rf)}(\mathbf{s})=\tilde{\mathbf{x}}}(\mathbf{w} \mid \theta, \alpha)},1 \right\} \\
    &= \min \left\{ \frac{f_{W(\bt) \mid W(\mathbf{s})=\tilde{\mathbf{x}} } (\mathbf{w}\mid \theta, \alpha) \cdot \frac{1}{c_{\rf}(\alpha)} C(\tilde{\mathbf{x}})  {\rf}^\alpha (\mathbf{w}^\prime ) f_{W(\bt) \mid W(\mathbf{s})=\tilde{\mathbf{x}} } (\mathbf{w}^\prime \mid \theta, \alpha)}{f_{W(\bt) \mid W(\mathbf{s})=\tilde{\mathbf{x}} } (\mathbf{w}^\prime\mid \theta, \alpha)\cdot \frac{1}{c_{\rf}(\alpha)} C(\tilde{\mathbf{x}})  {\rf}^\alpha (\mathbf{w}) f_{W(\bt) \mid W(\mathbf{s})=\tilde{\mathbf{x}} } (\mathbf{w} \mid \theta, \alpha)},1 \right\} \\
     &= \min \left\{ \frac{{\rf}^\alpha (\mathbf{w}^\prime ) }{  {\rf}^\alpha (\mathbf{w}) },1 \right\} 
\end{align*}
as acceptance rate for the proposal $\mathbf{w}^\prime$ .  The following theorem holds. 

\begin{theorem}
If $X$ is an $\rf$-Pareto process, the sample chain of the MCMC algorithm $\left(\mathbf{w}^{(n)} \right)_{n \in \mathbb{N}}$ converge to the conditional density in total variation norm, i.e.,
\begin{align*}
\lim_{n \to \infty}\left\|  \PP\left( \mathbf{w}^{(n)} \in \, \cdot  \right) - \PP\left( W^{(\rf)}(\bt) \in \, \cdot \,\middle \vert \,   W^{(\rf)}(\bs)=\bxtilde \right) \right\|_{\text{TV}} = 0 , 
\end{align*}
for arbitrary starting values $\mathbf{w}^{(0)}\in \operatorname{supp}\left(f_{W(\bt) \mid W(\mathbf{s})=\tilde{\mathbf{x}} } (\cdot \mid \theta, \alpha)\right)$  .
\end{theorem}   
For each of the $m$ independent observations $(x_1(s_0),x_1(\bs)), \dots , (x_m(s_0),x_{m}(\bs))$ of $(X(s_0),X(\bs))$ we repeat this procedure. So the MCMC algorithm produces conditional samples 
$$ \mathbf{w}_i \sim W^{(\rf)}(\bt) \, \vert \,   W^{(\rf)}(\bs)=x_i(\bs)/x_i(s_0) ,\, i \in 1,\dots,m.
$$
Under assumption \eqref{LimitLawForInference3} we know that $   X(\bt)=_d  W^{(\rf)}(\bt) \cdot X(s_0)$ holds. For real applications, we assume it to hold if we have a threshold exceedance, i.e.,  $$\rf(X(\bt))>u.$$
Now that we have fine grid simulations based on the observations, we can check if those are risk functional exceedances, i.e., so we check if
$$
\rf(\mathbf{w}_i\cdot x_i(s_0)) > u 
$$
holds. Therefore the simulations multiplied with the observations at $s_0$ satisfy
$$
\mathbf{w}_i\cdot x_i(s_0) \sim X(\bt) \, \vert \, \left\{ (X(s_0),X(\bs))=(x_i(s_0),x_i(\bs) ) , \rf(X(\bt))>u \right\}
$$
We can finally again define the set of risk functional exceedances over the threshold $u$ as 
\begin{align}
I(u):=\left\{ i\in \left\{ 1, \dots, m \right\}  : \, \rf\left(\mathbf{w}_i\cdot x_i(s_0) \right)>u \right\}.
    \label{set_of_exceedances_simulated}
\end{align}

\subsection{Estimation of Model Parameters $\theta$ and $\alpha$}
\label{subsec:unobservable-theta}  
Now that we again know the set of exceedances $I(u)$, we can proceed analogously to Section \ref{sec:risk-observable} for statistical inference. We use the same parametrization of the distribution of $W$  with parameter
$\theta \in \Theta \subset \mathbb{R}^k$ and a prior density $\pi(\theta,\alpha)$ on the parameter space $\Theta \times (0,\infty)$. {This allows us to condition} on an observation $(x(s_0),x(\bs))$ that belongs to an $\rf$-exceedance over the threshold $u$, i.e., $r(x)>u$. {Remind that the parameters $\theta,\alpha$ model only the behaviour conditional on $\rf(X) >u$ and, consequently, the actual probability of $\rf(X) > u$ does not depend on the parameters. Therefore, we obtain the joint posterior probability}

\begin{align*}
&f(\theta,\alpha \mid X(s_0) = x(s_0), \dots, X(s_n) = x(s_n), \rf(X)>u) \\
&\propto \pi(\theta,\alpha) 
    \cdot f_{(X(s_0)/u, X(\bs)/u)}\left( \frac{x(s_0)}{u}, \frac{x(\bs)}{u} \, \middle \vert \, \theta, \alpha, \rf(X) > u\right) \\
& \approx \pi(\theta,\alpha) \cdot f_{(P^{(\rf)}(s_0), P^{(\rf)}(\bs))} \left( \frac{x(s_0)}{u}, \frac{x(\bs)}{u} \, \middle \vert \, \theta, \alpha \right) \\
&\propto \pi(\theta,\alpha) \cdot  f_{(P_\alpha/\rf(W^{(\rf)}), W^{(\rf)}(\bs))} \left( \frac{x(s_0)}{u}, \frac{x(\bs)}{x(s_0)}  \, \middle \vert  \, \theta, \alpha \right) \\
&= \pi(\theta,\alpha) \cdot \int_{(0,\infty)^N}  f_{\left(P_\alpha/\rf\left(W^{(\rf)}(\bs,\bt)\right) \, \middle \vert  \, W^{(\rf)}(\bs,\bt)\right)}\left(\frac{x(s_0)}{u} \, \middle \vert \,  \frac{x(\bs)}{ x(s_0)},\frac{x(\bt)}{ x(s_0)} , \theta, \alpha\right) \\
& \qquad \qquad  \qquad \cdot f_{W^{(\rf)}(\bs,\bt)}\left(\frac{x(\bs)}{ x(s_0)} , \frac{x(\bt)}{ x(s_0)} \,\middle \vert \, \theta,\alpha \right)  \dd x(\bt)\\
&= \pi(\theta,\alpha) \cdot \int_{(0,\infty)^N}  \one\left\{x(s_0)\rf\left(\frac{x(\bs,\bt)}{x(s_0)}\right)>u\right\} \rf\left(\frac{x(\bs,\bt)}{x(s_0)}\right)^{-\alpha} \alpha \left(\frac{x(s_0)}{u}\right)^{-\alpha-1}  \\
& \qquad \qquad \qquad \cdot \frac{1}{c_{\rf}(\alpha,\theta)} \rf\left(\frac{x(\bs,\bt)}{x(s_0)}\right)^\alpha f_{W(\bs,\bt)}\left(\frac{x(\bs,\bt)}{x(s_0)} \, \middle \vert \,  \theta,\alpha\right)  \dd x(\bt)\\
&= \pi(\theta,\alpha) \cdot \alpha \left(\frac{x(s_0)}{u}\right)^{-\alpha-1}  \frac{1}{c_{\rf}(\alpha,\theta)}   \int_{(0,\infty)^N} \one\left\{x(s_0)\rf\left(\frac{x(\bs,\bt)}{x(s_0)}\right)>u\right\} f_{W(\bs,\bt)}\left(\frac{x(\bs,\bt)}{x(s_0)} \, \middle \vert \,  \theta,\alpha\right)  \dd x(\bt) \\
&= \pi(\theta,\alpha) \cdot \alpha \left(\frac{x(s_0)}{u}\right)^{-\alpha-1} \frac{1}{c_{\rf}(\alpha,\theta)}   f_{W(\bs)}\left(\frac{x(\bs)}{x(s_0)} \, \middle \vert \,  \theta,\alpha\right)  \\
& \qquad \qquad \qquad \qquad \cdot \PP \left( x(s_0) \rf(W(\bs))>u \, \middle \vert \, W(\bs)=\frac{x(\bs)}{x(s_0)},\theta,\alpha \right)
\end{align*}

Using all risk functional exceedances in $I(u)$ we get the posterior distribution
\begin{align*}
&f\left(\theta,\alpha \, \middle \vert \, \left\{ (x_i(s_0),x_i(\bs)) \right\}_{i\in I(u)} \right) \\
&\propto \pi(\theta,\alpha) \cdot  \prod_{i\in I(u)}  f_{X/u}\left( \frac{x_i(s_0)}{u}, \frac{x_i(\bs)}{u} \, \middle \vert \, \theta, \alpha\right) \\
&= \pi(\theta,\alpha) \cdot \prod_{i\in I(u)}  \alpha \left(\frac{x_i(s_0)}{u}\right)^{-\alpha-1} \frac{1}{c_{\rf}(\alpha,\theta)}   f_{W(\bs)}\left(\frac{x_i(\bs)}{x_i(s_0)} \, \middle \vert \,  \theta,\alpha\right) \\
& \qquad \qquad \qquad \qquad \cdot \PP \left( x_i(s_0) \rf(W(\bs))>u \, \middle \vert \, W(\bs)=\frac{x_i(\bs)}{x_i(s_0)},\theta,\alpha \right) \\
&=: \pi(\theta,\alpha) \cdot \tilde{g}\left(\left\{x_i(s_0), x_i(\bs) \right\}_{i \in I(u)} \, \middle \vert \,\theta,\alpha\right). 
\end{align*}
We again get a Metropolis Hastings algorithm to sample from this posterior distribution. Using a suitable proposal distribution $q((\theta,\alpha),(\cdot,\cdot))$, we accept the proposal $(\theta^\prime,\alpha^\prime)$ with probability 
\begin{align}
    a((\theta,\alpha),(\theta^\prime,\alpha^\prime))
    =\min\left\{ \frac{ q((\theta^\prime,\alpha^\prime),(\theta,\alpha))  \pi (\theta^\prime,\alpha^\prime)  \tilde{g}\left(\left\{x_i(s_0), x_i(\bs) \right\}_{i \in I(u)} \, \middle \vert \,\theta^\prime,\alpha^\prime\right)}{q((\theta,\alpha),(\theta^\prime,\alpha^\prime)) \pi (\theta,\alpha)  \tilde{g}\left(\left\{x_i(s_0), x_i(\bs) \right\}_{i \in I(u)} \, \middle \vert \,\theta,\alpha\right) },1 \right\}
    \label{acceptance rate}
\end{align}

\begin{theorem}
Let $X$ be an $\rf$-Pareto process and the proposal is chosen in such a way that irreducibility and aperiodicity are guaranteed. Further, let the set of exceedances $I(u)$ be fixed. Then, the output chain of the MCMC algorithm $\left(\theta^{(n)},\alpha^{(n)} \right)_{n \in \mathbb{N}}$ converges to the posterior in total variation norm, i.e.,
\begin{align*}
\lim_{n \to \infty}\left\|  \PP\left( (\theta^{(n)},\alpha^{(n)}) \in \, \cdot  \right) - \PP\left( (\theta,\alpha) \in \, \cdot \,\middle \vert \, \left\{x_i(s_0), x_i(\bs) \right\}_{i \in I(u)}\right) \right\|_{\text{TV}} = 0 , 
\end{align*}
for arbitrary starting values $\theta^{(0)} \in \Theta,\alpha^{(0)} > 0$.
\end{theorem}   

\begin{remark}
One advantage of the conditional simulation approach is that it {enables handling of} missing data without further changes. {In case of observations with missing values,} less points in the conditional simulation {are used} and the procedure for each observation {is changed} to conditional simulation on the sites $\tilde{\bs}_i:={\bs} \setminus \left\{ s_k : x_i(s_k) \text{ is missing} \right\}$. This can flexibly be done for each observation $x_1,\dots, x_m $.
\end{remark}

\section{Implementation} \label{sec:implementation}

\subsection{General Algorithm}

The whole procedure from Section \ref{sec:risk-unobservable} results in a nested two-step MCMC algorithm. For fixed parameters, Algorithm $\ref{sec:implementation}.\ref{alg:condX}$ implements the conditional fine grid simulation of $X$ from Subsection \ref{subsec:cond fine grid sampling} provided that an algorithm for the conditional simulation of $W$ is available. After some burn-in, we make use of the samples to obtain the risk functional evaluation on the fine grid and therefore are able to determine the threshold exceedances. 

\newpage

\begin{algo}
\label{alg:condX}
Cond-$X$ Algorithm.
\begin{tabbing}
\enspace Input: \enspace $\text{observations} \left( x_i(s_0), \dots, x_i(s_M) \right)_{i=1, \dots, \nobs} , \text{risk functional } \rf,$ \\
\enspace \phantom{Input:} \enspace$ \text{fine grid points } t_0, \dots, t_N, \text{parametric model for } W \text{ with parameters }(\theta,\alpha),$ \\
\enspace \phantom{Input:} \enspace  $\text{chain length } n_{\text{condX}}  $ \\ 
\enspace Output: \enspace  $\text{sample } \rf(X_i) \mid  X_i(s_0)=x_i(s_0), \dots, X_i(s_M)=x_i(s_M) \text{ under parameter }(\theta,\alpha)$ \\
\enspace Begin \\
 \qquad  For $i \in \left\{ 1, \dots, \nobs \right\} $ \\ 
 \qquad \qquad Initialize $w \sim W \mid W(s_1)=\frac{x_i (s_1)}{x_i (s_0)}, \dots, W(s_M)=\frac{x_i (s_M)}{x_i (s_0)} \text{ under parameter }(\theta,\alpha) $ \\
 \qquad \qquad \qquad  For $n \in \left\{ 1, \dots, n_{\text{condX}} \right\} $ \\
 \qquad \qquad \qquad \qquad Sample $w_{\text{prop}} \sim W \mid W(s_1)=\frac{x_i (s_1)}{x_i (s_0)}, \dots, W(s_M)=\frac{x_i (s_M)}{x_i (s_0)}$\\
 \qquad \qquad \qquad \qquad $ \text{under parameter }(\theta,\alpha)$ \\
 \qquad \qquad \qquad \qquad Sample $v\sim \text{Unif}[0,1]$ \\
 \qquad \qquad \qquad \qquad Set $a(w, w_{\text{prop}})=\min\left\{\frac{\rf^\alpha(w_{\text{prop}})}{\rf^\alpha(w)} , 1 \right\}$ \\
 \qquad \qquad \qquad \qquad If $v<a(w, w_{\text{prop}})$ \\
 \qquad \qquad \qquad \qquad \qquad Set $w=w_{\text{prop}}$ \\
 \qquad \qquad Set $w_i=w$ \\
 \qquad Return $\rf(X_i)=x_i(s_0)\cdot \rf(w_i)$ \\
\enspace End 
\end{tabbing}
\end{algo}

In the outer MCMC loop we estimate the parameters $\theta,\alpha$ {provided that the risk functional can be evaluated}. Algorithm $\ref{alg:MCMC}$ follows Subsection \ref{subsec:unobservable-theta}. After each parameter update step, Algorithm~$\ref{alg:condX}$ {is used} again to determine the threshold exceedances.  

\begin{algo}
\label{alg:MCMC}
MCMC Algorithm.
\begin{tabbing}
   \enspace Input: \enspace $\text{observations} \left( x_i(s_0), \dots, x_i(s_M) \right)_{i=1, \dots, \nobs} , \text{risk functional } \rf, \text{threshold } u$, \\
   \enspace \phantom{Input:} \enspace$\text{parametric model},\text{prior } \pi, \text{ proposal } q(\cdot,\cdot), \text{ chain length } n_{\text{MCMC}},$ \\
   \enspace \phantom{Input:} \enspace $\text{fine grid points } t_0, \dots, t_N  $ \\
   \enspace Output: \enspace  $ \text{Markov chain of parameters of length } n_{\text{MCMC}}$  \\
   \enspace Begin \\
   \qquad  Initialize $ \text{parameter } (\theta_0,\alpha_0) \text{ from prior } \pi$ \\
   \qquad for $n \in \left\{ 1, \dots, n_{\text{MCMC}} \right\} $ \\
   \qquad \qquad  Sample $\rf(X_i) \mid  X_i(s_0)=x_i(s_0), \dots, X_i(s_M)=x_i(s_M) \text{ under parameter }(\theta_{n-1},\alpha_{n-1})$ \\
   \qquad \qquad  $ \text{using Algorithm } \ref{alg:condX}$ \\
   \qquad \qquad  Set $I= \left\{ i \mid \rf(X_i)>u \right\}$ \\
   \qquad \qquad  Sample $( \theta_{\text{prop}},\alpha_{\text{prop}}) \sim q((\theta_{n-1},\alpha_{n-1}), \cdot)$ \\
   \qquad \qquad  Sample $v\sim \text{Unif}[0,1]$ \\
    \qquad \qquad  If $v<a((\theta_{n-1},\alpha_{n-1}),( \theta_{\text{prop}},\alpha_{\text{prop}}))$ \\
    \qquad \qquad   \qquad   Set $(\theta_n,\alpha_n)=(\theta_{\text{prop}},\alpha_{\text{prop}})$ \\
    \qquad \qquad Else \\
    \qquad \qquad  \qquad  Set $(\theta_n,\alpha_n)=(\theta_{n-1},\alpha_{n-1})$ \\
    \qquad  Return $(\theta_0,\alpha_0), \dots, (\theta_{n_{\text{MCMC}}},\alpha_{n_{\text{MCMC}}})$ \\
\enspace End 
\end{tabbing}
\end{algo}

The corresponding acceptance rate in Algorithm \ref{alg:MCMC}  depends on the choice of the spectral process and is specified in the next subsection.

\subsection{Brown-Resnick $\rf$-Pareto process}

{Being one of the most popular process models in spatial extremes, we consider} the Brown-Resnick Pareto process introduced in Subsection \ref{subsec:BR} {in more detail. For this example,} the spectral process $W$ is given as the log-Gaussian process
$$
 W(s) = \exp\left( \frac 1 \alpha \left[G(s)-G(s_0) - \gamma(s-s_0) \right]\right), \quad s \in \RR^d,
$$
parameterized with the power variogram given in (\ref{def:powervariogram}). Since the Brown-Resnick process only depends on the semi-variogram of the underlying Gaussian process $G$, we can choose any $G$ with that semi-variogram. Here, we choose $G$ as a centred Gaussian process with covariance
$$
\Cov(G(s),G(t)):= c \left( \|s\|^\beta + \|t\|^\beta - \|s-t\|^\beta\right),
$$
the so-called \textit{fractional Brownian field}. With this specific process and a regular grid structure, we have access to a very fast simulation algorithm called \textit{circulant embedding}. One makes use of the fact that $n\times n$ circular matrices can be diagonalized via the Fast Fourier Transform (FFT) with a complexity of only $\mathcal{O}(n \log n) $. First the process $G$ is modified to have a stationary covariance structure. Then the covariance matrix is embedded in a block circulant matrix and diagonalized via the FFT allowing for fast samples from the resulting process which then can be transformed back to samples from $G$, see {Section 4 in \cite{kroese-botev-2015}}. 

\begin{algo}
\label{alg:Gauss}
Log-Gaussian Simulation Algorithm.
\begin{tabbing}
   \enspace Input: \enspace $ \text{fine grid points } t_0, \dots, t_N, \text{parametric Brown-Resnick model with parameter }(\theta,\alpha) $\\
   \enspace Output: \enspace  $\text{sample from } W  \text{ under parameter }(\theta,\alpha) \text{ on fine grid}$  \\
   \enspace Begin \\
   \qquad Sample $ g(t_1),\dots, g(t_N) \sim \text{fractional Brownian field} \text{ under parameter } \theta = (c,\beta)  $ \\
   \qquad  Return $w_j=\exp\left(\frac{1}{\alpha}(g(t_j)-g(s_0)-\gamma_\theta (t_j,s_0))\right)$ \\
   \enspace End 
\end{tabbing}
\end{algo}

The corresponding acceptance rate in Algorithm $\ref{alg:MCMC}$ contains the posterior 
\begin{align*}
  f(\theta,\alpha \mid x(s_0), \dots, x(s_n)) = \pi(\theta,\alpha) \cdot \alpha \left(\frac{x(s_0)}{u}\right)^{-\alpha-1} \frac{1}{c_{\rf}(\alpha,\theta)}   f_{W(\bs)}\left(\frac{x(\bs)}{x(s_0)} \, \middle \vert \,  \theta,\alpha\right)  
\end{align*}
defined in (\ref{acceptance rate}), where the estimation of the reciprocal normalizing constant $1/c_{\rf}(\alpha, \theta)$ is done
via fractional Brownian field simulations, see Subsection \ref{subsec:estimation of 1/c}, and the likelihood of the spectral process $f_{W(\bs)}\left(x(\bs) / x(s_0) \, \middle \vert \,  \theta,\alpha\right) $ is a log-Gaussian density.

\newpage

\begin{algo}
\label{alg:acceptance rate}
Acceptance Rate MCMC Algorithm.
\begin{tabbing}
   \enspace Input: \enspace $\text{exceedance observations} \left( x_i(s_0), \dots, x_i(s_M) \right)_{i\in I_{\text{exceed}}} , \text{risk functional } \rf, $ \\
   \enspace \phantom{Input:} \enspace $\text{prior } \pi, \text{ proposal } q(\cdot,\cdot), \text{ parametric model with parameter } (\theta_{\text{prop}},\alpha_{\text{prop}}) \text{ and } $ \\
   \enspace \phantom{Input:} \enspace  $ (\theta_{\text{old}},\alpha_{\text{old}}) , \text{sample size } n_{\text{CondGauss}}, \text{ sample size } n_{\text{Gauss}}, \text{ fine grid points } t_0, \dots, t_N  $ \\
   \enspace Output: \enspace  $\text{acceptance rate } a((\theta_{\text{old}},\alpha_{\text{old}}), (\theta_{\text{prop}},\alpha_{\text{prop}}))  $  \\
   \enspace Begin \\
   \qquad For $\theta\in \left\{(\theta_{\text{prop}},\alpha_{\text{prop}}), (\theta_{\text{old}},\alpha_{\text{old}})\right\}$ \\
    \qquad \qquad For $i \in I_{\text{exceed}}$ \\
   \qquad \qquad \qquad Set {$ a= f_{W(\bs)}\left(x(\bs) / x(s_0) \, \middle \vert \,  \theta,\alpha\right)$ (log-Gaussian density)}\\
   \qquad \qquad \qquad Set $  b=  \alpha \left(\frac{x_i(s_0)}{u}\right)^{-\alpha-1}$ \\
    \qquad \qquad \qquad  Estimate $c= \EE(\rf(W)^\alpha)$ \text{ by } $n_{\text{Gauss}}$ {\text{ runs of Algorithm}~\ref{alg:Gauss} \text{ with parameter }$(\theta,\alpha)$}  \\
    \qquad \qquad \qquad {Estimate $d=\PP \left( x_i(s_0) \rf(W(\bs))>u \, \middle \vert \, W(\bs)=x_i(\bs) / x_i(s_0),\theta,\alpha \right)$ by $n_{\text{CondGauss}}$} \\
    \qquad \qquad \qquad {simulations of log-Gaussian process $W \mid W(\bs)= x_i (\bs) / x_i (s_0)$} \\
    \qquad \qquad \qquad  Set $\text{log likelihood}((\theta,\alpha),i)=\log a + \log b - \log c {+ \log d}$ \\
    \qquad \qquad Set $\text{log likelihood}(\theta,\alpha)=\sum_{i \in I_{\text{exceed}}}\text{log likelihood}((\theta,\alpha),i)$ \\
   \qquad  Return $\min\left\{1,\frac{\pi((\theta_{\text{prop}},\alpha_{\text{prop}})) q((\theta_{\text{prop}},\alpha_{\text{prop}}),(\theta_{\text{old}},\alpha_{\text{old}})) \text{ likelihood}((\theta_{\text{prop}},\alpha_{\text{prop}}))}{\pi((\theta_{\text{old}},\alpha_{\text{old}})) q((\theta_{\text{old}},\alpha_{\text{old}}),(\theta_{\text{prop}},\alpha_{\text{prop}})) \text{ likelihood}((\theta_{\text{old}},\alpha_{\text{old}}))}\right\} $\\
   \enspace End 
\end{tabbing}
\end{algo}

\subsection{Estimation of $1/c_r$}
\label{subsec:estimation of 1/c}

Algorithm \ref{alg:acceptance rate} requires two estimation steps. While the probabilities
$\PP( x_i(s_0) \rf(W(\bs))>u \, \vert \, W(\bs)= x_i(\bs) / x_i(s_0),\theta,\alpha)$ allow for straightforward unbiased estimation, estimation of the reciprocals of the constants $c_\rf (\theta,\alpha)$ is more involved. The latter enter the acceptance rate ad the difference between the two log-likelihoods of the constants $c_\rf (\theta,\alpha)$ for two different parameters, i.e., $\log(c_\rf (\theta_{\text{old}},\alpha_{\text{old}}))-\log(c_\rf (\theta_{\text{prop}},\alpha_{\text{prop}}))$.
For the variance of the difference of two plug-in estimators $\log(\hat c_1)$ and $\log(\hat c_2)$, it holds
\begin{align*}
    &\Var(\log(\hat c_1)-\log(\hat c_2)) \\
    =&\Var(\log(\hat c_1))+\Var(\log(\hat c_2))-2 \surd{\Var(\log(\hat c_1))} \surd{\Var(\log(\hat c_2))} \Corr (\log(\hat c_1),\log(\hat c_2)).
\end{align*}
Therefore the variance of the difference will get small if the variances of the single estimators are small and the correlation between them is large. 
To estimate $\log(c_r(\theta,\alpha))=\log(\EE\left[ \left( r(W_{\theta,\alpha})^\alpha\right) \right] )$ the estimator $\log(\Bar{c}_N(\theta,\alpha)):=\log\left( 1 / N \sum_{i=1}^N \rf(w_i)^\alpha\right)$, where $w_i$ are independent samples from $W_{\theta,\alpha}$ {as defined in \eqref{def:BR} via} $$
 W_{\theta,\alpha}(s) = \exp\left( \frac 1 \alpha \left[G(s)-G(s_0) - \gamma(s-s_0) \right]\right), \quad s \in \RR^d,
$$
where $G_\theta$ is a centred Gaussian process with stationary increments and semi-variogram depending on $\theta$. From $N$ repetitions of $d$-dimensional Gaussian noise, one can build two correlated samples of $W_{\theta_{\text{old}},\alpha_{\text{old}}}$ $W_{\theta_{\text{prop}},\alpha_{\text{prop}}}$ and with the estimators $\hat c_1=\Bar{c}_N(\theta_{\text{prop}},\alpha_{\text{prop}})$ and $\hat c_2=\Bar{c}_N(\theta_{\text{prop}},\alpha_{\text{prop}})$
accordingly. 

{To ensure that the variance of estimator $ \Bar{c}_N:= {n^{-1}} \sum_{i=1}^N \rf(w_i)^\alpha$ is sufficiently small, where $w_i$ are independent samples from $W$, of $c_r(\alpha,\theta)=\EE\left[ \left( r(W)^\alpha\right) \right]$, 
we propose} a dynamic approach. {Using the classical central limit and the delta method, we obtain}
\begin{align*}
    \surd{N}\left({\log\Bar{c}_N-\log{c}}\right)\to \mathcal{N}\left(0,\frac{1}{{c^2}}\Var(\rf(W)^\alpha)\right).
\end{align*}
This leads to a heuristic for the variance of the estimator {which is} 
\begin{align*}
    \Var\left({\log\Bar{c}_N}\right) \approx \frac{1}{N} \frac{1}{{c^2}}\Var(\rf(W)^\alpha)
\end{align*}
We use the above heuristic to guarantee that the {standard deviation of the plug-in estimator} is smaller than a certain accuracy $q$ {by choosing $N$ sufficiently large such} that 
\begin{align}
  \frac{1}{c}\left\{ \frac{1}{N} \Var(\rf(W)^\alpha) \right\}^{1/2} < q
  \label{N_est_c estimation inequality}
\end{align}
holds. In practice, {estimation is started} with a relatively low $N$ {which then might be increased until the inequality \eqref{N_est_c estimation inequality} holds or an upper bound is reached}.

\section{Numerical results}
In numerical experiments, we consider the Brown-Resnick $\rf$-Pareto process with the risk functional $r$ being equal to the mean on the finer set of sites $\mathbf{t}:= ( t_1, \dots, t_N)$, i.e.\, 
$$
\rf(f)=\rf(f(\mathbf{t}))= \frac{1}{N} \sum_{k=1}^{N} f(t_k).
$$
In order to obtain samples from that Brown-Resnick $\rf$-Pareto process, we use an MCMC approach as {suggested} in \cite{dombry-ribatet-2015}. We {generate $m$ independent sample} and only keep the values at the coarse sites $s_0,s_1,\dots,s_n$, i.e., {we consider $m$ realizations} 
$$
x_i=(x_i(s_0),x_i(s_1),\dots,x_i(s_n)), \quad  i= 1,\dots,m
$$
of the Brown-Resnick $\rf$-Pareto process with extreme value index {$1/\alpha=1/2$} and power variogram $\gamma(h)=c \|h\|^\beta$, with {$\beta=0.5$ and $c=3$}. 
\begin{figure}[htbp]
    \centering
    \includegraphics[width=0.7\textwidth]{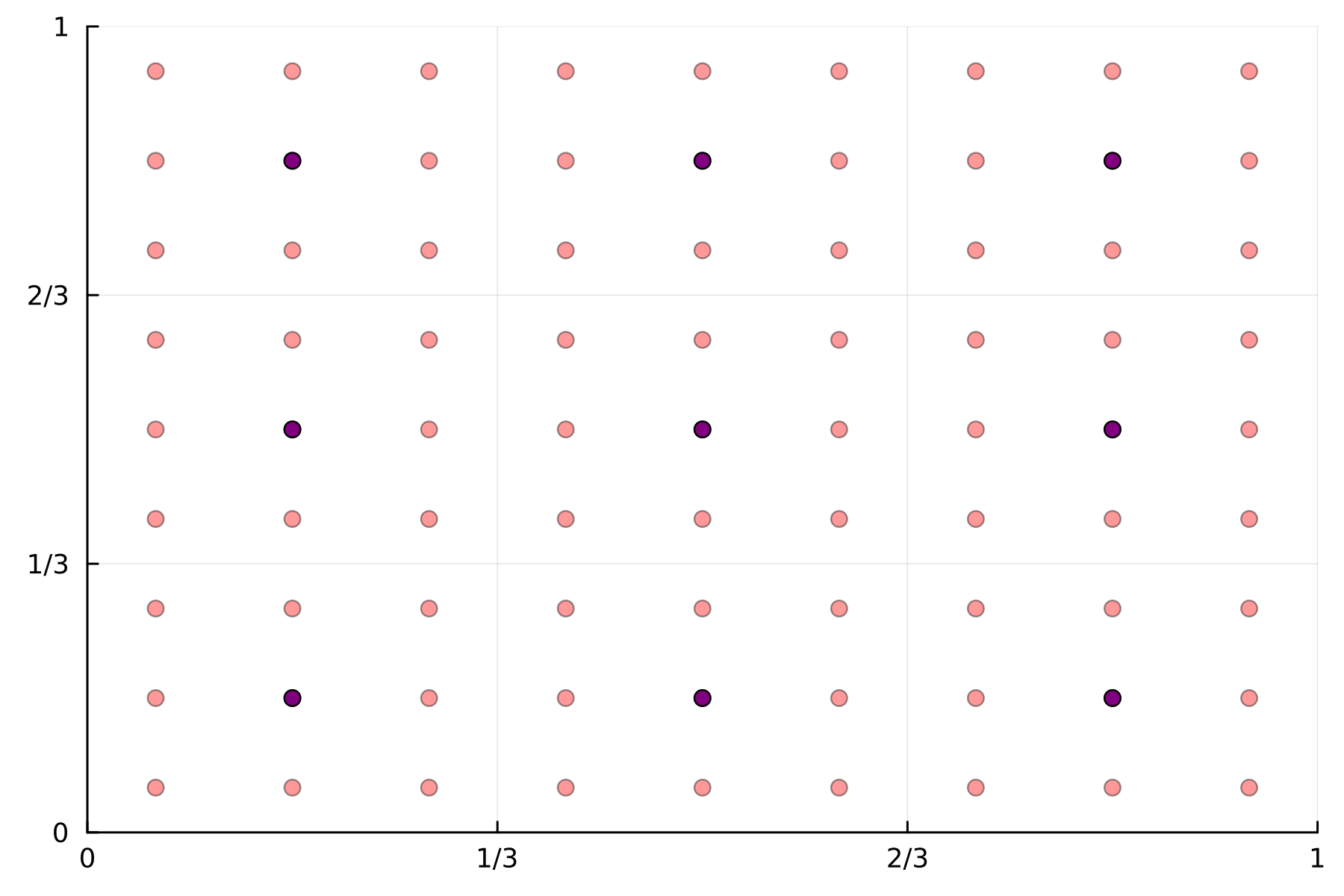}
    \caption{Finer sites for risk functional $\rf$ evaluation in red and coarse grid as observation points in purple.}
    \label{fig:stations-grid}
\end{figure}

{The sampling design is chosen to be} a nine-by-nine grid of $N=81$ finer sites and $n+1=9$ coarse observations, including one normalizing site $s_0$, {as displayed in} Figure \ref{fig:stations-grid}). We compare the conditional sampling approach in Section \ref{sec:risk-unobservable} with the state-of-the-art approach in Section \ref{sec:risk-observable} of using an approximate observable risk functional 
$$
\rf_{\text{approx}}(f):=\rf_{\text{approx}}(f(\bs))= \frac{1}{n} \sum_{k=1}^{n} f(t_k).
$$
being the mean over the observed sites. {This choice is natural due to the symmetry of the sampling design and can be further justified by some preliminary considerations showing that this choice of $\rf_{\text{approx}}$ corresponds to the linear binary classifier with the minimal extremal risk as defined in \cite{LNO-2025}.}

{More precisely, we} sample {$m=100$} observations, choose the threshold $u=1$, a chain length of $10000$ and discard the first $1000$ realizations as burn in. {For both $\log \alpha$ and $\log c$, we consider normal random walk proposals and zero mean normal priors.}
For $\beta \in (0,2)$, we {choose a uniform proposal an interval of width 0.2 centred} around the current value, fixing it at the margins to avoid values out of $(0,2)$, {and a prior that is} uniform on $(0,2)$. For the starting values of the parameters, we sample from the priors and then let the one-step-approximate algorithm run for $1000$ steps and take that sample as staring value for both algorithms. We then used the posterior mean and posterior median of the remaining samples as a point estimator for the parameters. We repeat the whole procedure 100 times and {display the root mean squared errors (RMSE)} of the estimators in Table \ref{tab:RMSE} {showing} significant improvements in the RMSE for both the mean and the median estimators of $c$ and $\alpha$ ranging from {$10$ to $15$} percent. Only in the estimation of the parameter $\beta$ there is no significant difference between the two methods. We believe that the gain in efficiency comes from the ability of the conditional simulation to better capture the complexity in the risk functional and the corresponding $r$-Pareto process.

\begin{table}[h]
\centering
\begin{tabular}{c|ccc|ccc}
       & \multicolumn{3}{c|}{\text{Median}} & \multicolumn{3}{c}{\text{Mean}} \\
      Method & $\beta$ & $c$ & $\alpha$ & $\beta$ & $c$ & $\alpha$ \\
\hline
conditional  &        0.059  &  0.761   &    0.279 & 0.058    &  0.698  &  0.265     \\
approx  &      0.056  &  0.868   &    0.309    &    0.056      &   0.813  &     0.309  \\
\end{tabular}
\caption{{Root mean squared errors from simulation study based on 100 repetitions.}}
\label{tab:RMSE}
\end{table}

\section{Discussion}

We present a full Bayesian approach for inference of $r$-Pareto processes for arbitrary partially unobservable risk functionals $r$. In an alternating two-step MCMC algorithm we first sample {the underlying process} on a high resolution grid conditional on the coarse observation data to {obtain a classification of extremes via} risk functional exceedances. Secondly, the exceedances {are used to update the} parameters in a Metropolis-Hastings step.
{Conditional sampling allows for the evaluation of the risk functional at an arbitrarily high resolution and does not require an approximation of the functional of interest from sparse data}. This leads to a {remarkable} accuracy gain in the estimation of dependence parameters---{in the numerical experiments, we see} up to  {$15$} percent improvement in the RMSE. We believe that for more complex risk functionals and rougher processes, the accuracy gain in the estimation might be even {larger}. 

In statistical methods and their applications missing data tend to be a challenge. Through the conditional sampling approach, missing observations can be tackled naturally {by reducing the number of conditions without any further modification of the method.}

The Bayesian framework opens up the {full range of instruments from Bayesian inference}. From the posterior distribution samples, we can directly calculate different point estimates, credible intervals, and quantiles of interest. In addition, we can easily sample from the fitted model using the full information of the posterior distribution {appropriately accounting for different types of uncertainties}. Further Bayesian techniques for model comparison, such as the Bayes factor or the posterior predictive distribution, could be used to compare different variograms in their goodness of fit.

{In the numerical experiments, We did not compare} different setups of the two-step algorithm. Different sample sizes and different combinations of the two steps might also lead to better and faster convergence of the chains. Another possible improvement would include more sophisticated MCMC approaches. So far, only standard log-Gaussian proposals and a Metropolis-Hastings-like algorithm {have been considered}.

\section*{Acknowledgement}

Funded by Deutsche Forschungsgemeinschaft (DFG, German Research Foundation) under Germany's Excellence Strategy -- EXC 2075 -- 390740016. We acknowledge the support by the Stuttgart Center for Simulation Science (SimTech).

\bibliographystyle{abbrvnat}
\bibliography{lit.bib}

\end{document}